
\input phyzzx
\input tables
\voffset = -0.4in
\footline={\ifnum\pageno=1 \nulline \else\newfootline \fi}
\def\nulline{{\hfill}}
\def\newfootline{\advance\pageno by -1\hss\tenrm\folio\hss}
\rightline {March 1994}
\rightline {QMW--TH--94/05}
\title {Background Symmetries In \break
Orbifolds With Discrete Wilson Lines.}
\author{ A. Love$^{a}$, W. A.
Sabra$^{a *}$\ and \ S. Thomas$^{b **}$}
\footnote*{e-mail address: UHAP012@VAX.RHBNC.AC.UK.}
\footnote{**}{e-mail address: S.THOMAS@QMW.AC.UK.}
\address {$^{a}$Department of Physics,\break
Royal Holloway and Bedford New College,\break
University of London,\break
Egham, Surrey, U.K.}
\address {$^{b}$
Department of Physics,\break
Queen Mary and Westfield College, \break
University of London,\break
Mile End Road, London,  U.K.}
\abstract {Target space symmetries are studied for orbifold compactified
string
theories
containing Wilson line background fields. The symmetries determined are for
those moduli which contribute to the string loop threshold corrections of the
gauge coupling constants. The
groups found are subgroups of the modular group $PSL(2, Z)$ and depend on the
choice of discrete Wilson lines and the shape of the
underlying six-dimensional lattice.}
\endpage
\REF\one{L. Dixon, J. A. Harvey, C. Vafa and E. Witten, {\it Nucl. Phys}.
{\bf B261} (1985) 678; {\bf B274} (1986) 285.}
\REF\two{ K. Narain, M. Sarmadi and C. Vafa, {\it Nucl. Phys}.
{\bf B288} (1987) 551.}
\REF\three{ L. E. Ibanez, H. P. Nilles and F. Quevedo, {\it Phys. Lett}.
{\bf B187} (1987) 25; {\it Phys. Lett}.
{\bf B192} (1987) 332. }
\REF\four{ A. Font, L. E. Ibanez,
F. Quevedo and A. Sierra, {\it Nucl. Phys}. {\bf B331} (1991) 421.}
\REF\five{R. Dijkgraaf, E. Verlinde and H. Verlinde, {\it Comm. Math.
Phys}.115
(1988) 649.}
\REF\six{ S. Ferrara, D. L\"{u}st, A. Shapere and S. Theisen, {\it Phys.
Lett}.
{\bf B225} (1989) 363.}
\REF\seven{ R. Dijkgraaf, E. Verlinde and H. Verlinde, On Moduli Spaces of
Conformal Field Theories with $c \geq 1$, Proceedings Copenhagen Conference,
Perspectives in String Theory,
edited by P. Di Vecchia and J. L. Petersen, World Scientific, Singapore,
1988.}
\REF\eight {L. J. Dixon, V. S. Kaplunovsky and J. Louis,  {\it Nucl. Phys}.
{\bf B355} (1991) 649.}
\REF\nine{V. S. Kaplunovsky, {\it Nucl. Phys}. {\bf B307} (1988) 145}
\REF\ten{J. P. Derenddinger, S. Ferrara, C. Kounas and F. Zwirner, {\it
Nucl.
Phys.} {\bf B372} (1992) 145, {\it Phys. Lett}. {\bf B271} (1991) 307.}
\REF\eleven {L. E. Ibanez, D. Lust and G. G. Ross,  {\it Phys. Lett}.
{\bf B272} (1991) 25.}
\REF\twelve{D. Bailin and A. Love,  {\it Phys. Lett}.
{\bf B278} (1992) 125; {\it Phys. Lett}.
{\bf B292} (1992)  315.}
\REF\thirteen{ A. Shapere and F. Wilczek, {\it Nucl. Phys}. {\bf B320} (1989)
669.}
\REF\fourteen{ M. Dine, P. Huet and N. Seiberg,
{\it Nucl. Phys}.  {\bf B322} (1989) 301.}
\REF\fifteen{ J. Lauer, J. Mas and H. P. Nilles, {\it Nucl. Phys}. {\bf B351}
(1991)
353.}
\REF\sixteen{ K. Kikkawa and M. Yamasaki, {\it Phys. Lett}. {\bf B149}, (1984)
357;
N. Sakai and I. Senda, {\it  Prog. Theor. Phys}. 75 (1984) 692}
\REF\seventeen{A. Giveon, E. Rabinovici and G. Veneziano,
{\it Nucl. Phys.} {\bf B322} (1989) 167.}
\REF\eighteen {W. Lerche, D. L\"ust and N. P. Warner,
{\it Phys. Lett}. {\bf B231} (1989) 417.}
\REF\nineteen {K. S. Narain, {\it Phys. Lett}. {\bf B169} (1987) 41.}
\REF\twenty{ K. S. Narain, M. H. Sarmadi  and E. Witten,   {\it Nucl. Phys.}
{\bf
B279} (1987) 369.}
\REF\twentyone{M. Spalinski, {\it Nucl. Phys}.
{\bf B377} (1992) 339.}
\REF\twentytwo{Y. Katsuki, Y. Kawamura, T. Kobayashi, N. Ohtsubo, Y. Ono and
K.
Tanioka, {\it Nucl. Phys}. {\bf B341} (1990) 611.}
\REF\twentythree{ P. Mayr and S. Stieberger, {\it Nucl. Phys}. {\bf B407}
(1993)
725.}
\REF\twentyfour{D. Bailin, A. Love, W. A. Sabra and S. Thomas,
{\it Phys. Lett.} {\bf  B320} (1994) 21.}
\REF\twentyfive{D. Bailin, A. Love, W. A. Sabra and S. Thomas, {\it Mod. Phys.
Lett}. {\bf A 9} (1994) 67.}
\REF\twentysix{ T. Kobayashi and  N. Ohtsubo, preprint DPKU--9103.}
\REF\twentyseven{ L. E. Ibanez, J. Mas, H. P. Nilles and F. Quevedo, {\it
Nucl.
Phys.}
{\bf B301} (1988) 157.}
\REF\twentyeight{L. E. Ibanez and D. L\"{u}st , {\it Nucl. Phys}. {\bf B382}
(1992)
305.}
\REF\twentynine{M. Spalinski,  {\it Phys.  Lett}. {\bf B275} (1992) 47;
J. Erler, D. Jungnickel and H. P. Nilles, {\it Phys. Lett}. {\bf B276} (1992)
303.}
\REF\thirty{J. Erler and M. Spalinski, preprint, MPI-PH-92-61,
TUM-TH-147-92}
\REF\thirtyone{ D. Bailin, A. Love, W. A. Sabra and S. Thomas, QMW-TH-93-31,
SUSX-TH-93-17, to appear in {\it Mod. Phys. Lett.} {\bf A}.}
\REF\thirtytwo{G. L. Cardoso, D. Lust and T. Mohaupt, HUB-IEP-94/6.}
\REF\thirtythree{T. Mohaupt, MS-TPI 93-09.}
\chapter{Introduction}
The compactified string background can be described by a two-dimensional
conformal field theory with a  central charge equal the number of the
compactified dimensions. Of all known conformal field theories, orbifold
models
offer  phenomenologically promising string compactified backgrounds
[\one-\four].  Orbifolds are characterized by a set of moduli which
parametrize, locally, the string background  and correspond to the truly
marginal deformations of the underlying conformal field theory [\five, \seven].
These moduli
appear as massless scalars in the low-energy effective action with flat
potentials to all orders in  string perturbation theory.
A fundamental difference between  compactified  string
theories and those of
Kaluza-Klein is the fact that the former theories have a novel symmetry
known as duality
[\five, \seven, \thirteen,\fourteen, \fifteen, \sixteen, \seventeen,
\eighteen].
This  symmetry generalizes the well known
$R\rightarrow {1/2R}$ symmetry for circle compactification with $R$ being the
radius of the circle. Target  space duality symmetry restricts the form of the
low-energy effective action and in particular any possible non-perturbative
superpotential for the moduli  [\six].

The duality group for two-dimensional toroidal compactification
is given by two copies of the modular group $PSL(2,Z)$ acting on the two
complex moduli, $T $ and $U$ describing the two-dimensional target space
[\seven] via the fractional linear transformations:
$$U\rightarrow {aU+b\over cU+d},\qquad
T\rightarrow {a'T+b'\over c'T+d'},$$
where $ad-bc=1$ and $a'd'-b'c'=1.$
Note that the $U$-symmetry is not of stringy origin and  is also a symmetry of
the Kaluza-Klein compactification.

For the two-dimensional ${\bf Z}_N$ orbifold, $(N\not=2)$, the $U$ modulus is
frozen, i.e.,  its value is fixed to a constant phase factor and the duality
group  associated with the complex $T$ modulus is always the modular group
$PSL(2,Z)$.
Clearly, in the realistic cases,  the modular group can be realized as a
duality group for each complex modulus associated with the three complex
planes
of the six-dimensional orbifold provided that the orbifold lattice is
decomposable in the following form ${\bf\Lambda}_6={\bf\Lambda}_2\bigoplus
{\bf\Lambda}_2\bigoplus {\bf\Lambda}_2.$

Whilst the  duality group associated with the complete spectrum of
orbifold models is rather complicated in general, a somewhat  simpler task
is to study the duality symmteries of the twisted sectors contributing to the
moduli dependent threshold
corrections. Moreover, a knowledge of the latter will be important
in any investigation of mechanisms that fix the values of the various moduli.
In ref. [\eight] the string-loop threshold corrections to the gauge coupling
constants [\eight, \nine, \ten, \eleven, \twelve, \twentyeight] arising from
the twisted sectors with invariant
planes of the orbifold were calculated. The models of [\eight] are such that
the six-dimensional lattice ${\bf\Lambda}_6$ of the orbifold is decomposable
into a
direct sum of a two-dimensional and a four-dimensional sub-lattices, $i. e.$,
${\bf\Lambda}_6={\bf\Lambda}_2\bigoplus{\bf\Lambda}_4$, with the
twist-invariant plane lying
in ${\bf\Lambda}_2$. These moduli-dependent threshold corrections turn out to
be
invariant under the modular group.
This is not surprising since the modular group leaves the spectrum of twisted
states contributing to the threshold corrections  invariant.
However there are many orbifold models with lattices that  do not admit the
above simplifying decomposition [\twentysix, \twentytwo]. For the  more general
choice of
lattices it was found  that the  duality group for the moduli appearing in the
threshold corrections to the gauge couplings [\twentythree, \twentyfour,
\twentyfive]  is, in
some cases, a congruent subgroup of the modular group. The congruent  groups
$\Gamma_0(n)$  and $\Gamma^0(n)$ are represented
by the following set of matrices
$$\pmatrix{a&b\cr c&d};\qquad  ad-bc=1,
$$ with
$c=0 \  \hbox{(mod} \ n)$
and $b=0 \  \hbox{(mod} \ n)$, respectively.

In the presence of discrete Wilson lines background, the duality group is
known
to be broken [\twentyeight-\thirtyone]. In [\thirtyone] the
symmetries of the moduli relevant to string loop threshold corrections in
orbifold theories with Wilson lines were considered with the assumption that
the orbifold lattice ${\bf\Lambda}_6={\bf\Lambda}_2\bigoplus{\bf
\Lambda}_4$, with the
invariant plane of the twist lying in ${\bf\Lambda}_2$.
It is our purpose to generalize the results of [\thirtyone] to include other
choices of the orbifold lattice.
We organize this work as follows. First we review the one-dimensional
compactifiaction, $i. e.,$ compactification of a string coordinate on the
circle.
It is also shown that discrete Wilson lines background can break the
$R\rightarrow {1/2R}$ duality symmetry. In section 2,  the
analysis is extended to the more realistic toroidal and orbifold
compactification in the absence of Wilson lines. In section 3  the duality
symmetry for the two-dimensional case is considered. This analysis will be
relevant to the study of duality symmetries of the twisted sectors
contributing to the threshold corrections in the six-dimensional case. In the
remaining sections, the duality symmetries for the
moduli contributing to the threshold correction in orbifold models with Wilson
lines background are presented.
It is found  that the duality group is in most cases given by a subgroup of the
modular group $PSL(2, Z)$. The constraints satisfied by the parameters of the
modular group depend  on the choice of the Wilson line
background and the shape of the underlying six dimensional lattice.
\chapter{Circular Compactification}
Consider a string coordinate compactified on a circle of radius $R$, the
worldsheet action for the compact coordinate is given by
$$S={1\over 2\pi}\int d\sigma d\tau\eta^{\alpha\beta}\partial_\alpha
X\partial_\beta X,\eqn\ac$$
where $(\sigma,\tau)$ and $\eta^{\alpha\beta}$ are the coordinates and the
metric of the worldsheet. If the string is parametrized by $\sigma\in [0,
2\pi]$, then periodicity on the circle implies
$$X(\sigma+2\pi)=X(\sigma)+2\pi nR\eqn\gif$$ where $n,$
the winding number, is an integer representing the number of times the string
wrap around the circle.

The compact coordinate $X$ satisfies a free wave equation with the boundary
condition \gif,  therefore its solution
splits into a left and right moving parts given by
$$\eqalign{X_R=&x_R-{1\over2} p_R(\sigma-\tau)+{i\over 2}
\sum_{k\not=0}{1\over
k}
\alpha_ke^{ik(\sigma-\tau)}\cr
X_L=&x_L-{1\over2} p_L(\sigma+\tau)+{i\over 2} \sum_{k\not=0}{1\over k}
\tilde\alpha_ke^{-ik(\sigma+\tau)},}\eqn\modes$$
where $\alpha_k$ and $\tilde\alpha_k$ are the oscillators, and the left and
right moving momenta are given by
\footnote* {for string slope parameter $\alpha'$ taken to be $1/2$}
$$p_L={m\over2R}+nR, \quad\quad p_R={m\over2R}-nR.\eqn\lrm$$
The $m$ dependent term is simply a consequence of quantum mechanics, while
that of $n$  is of stringy nature and is due to the fact that strings are
extended objects and can wrap around the circle.

The vertex operators of the conformal field theory of the compact coordinate
$X$ have the following  moduli-dependent scaling and spin
$$\eqalign{H=&{1\over2} p_L^2+{1\over2} p_R^2={m^2\over4R^2}+n^2R^2,\cr
S=&{1\over2} p_L^2-{1\over2} p_R^2=mn.}\eqn\sam$$
Clearly, the spectrum is invariant under the transformation
$$R\leftrightarrow {1\over2R}; \qquad m\leftrightarrow n.\eqn\dua$$
This is the simplest form of duality symmetry in compactified string theories.

Now consider the circular compactification in the presence of a discrete
Wilson
line background. In this case, the  left and right moving momenta take the
form
$$\eqalign{P_L=&\Big({(m-{1\over2}{\bf A}^t{\bf C}{\bf A}n-{\bf A}^t{\bf
C}{\bf
l})\over 2R} +Rn,\quad  {\bf l}+{\bf A}n\Big)\equiv \Big(p_L, \quad  {\tilde
p}_L\Big),\cr P_R=&\Big({(m-{1\over2}{\bf A}^t{\bf C}{\bf A}n-{\bf A}^t{\bf
C}{\bf l})\over 2R}-Rn, \quad {\bf 0}\Big)\equiv
\Big({p}_R,\quad  {\bf 0}\Big)}\eqn\wlrm$$
where ${\bf l}$ is the momentum on the $E_8\times E'_8$ lattice,  $\bf A$ is
the discrete Wilson line background fields and $\bf C$ is the
Cartan metric for the $E_8\times E'_8$ lattice.

In the presence of Wilson lines, the vertex operators have the following
moduli-dependent scaling and spin
$$\eqalign{H=&{(m-{1\over2}{\bf A}^t{\bf C}{\bf A}n-{\bf A}^t{\bf C}{\bf
l})^2\over4R^2}+n^2R^2+{1\over 2} {\tilde p}_L^t{\bf C} {\tilde p}_L\cr
S=&(m-{1\over2}{\bf A}^t{\bf C}{\bf A}n-{\bf A}^t{\bf C}{\bf l})n+
{1\over 2}{\tilde p}_L^t{\bf C}{\tilde p}_L,}\eqn\rk$$
where ${}^t$ denotes matrix transpose. Note that we have included the
$R$-independent term ${1\over 2} {\tilde p}_L^t{\bf C} {\tilde p}_L$
because  it
depends on the winding number.

We  look for duality symmetries that act only the
components of $p_L $ and $p_R $ associated with the orbicircle, leaving
the internal ${\tilde p}_L$ momentum and the discrete Wilson line $\bf A$
invariant.
An orbicircle of radius $R$ is constructed from a circle of the
same radius, by modding out by a (${\bf Z}_2 $-valued ) reflection symmetry.
Consider the following duality symmetry,
$$R\leftrightarrow {1\over 2\rho R}; \quad
(m-{1\over2}{\bf A}^t{\bf C}{\bf A}n-{\bf A}^t{\bf C}{\bf l})\leftrightarrow
{1\over \rho}n.\eqn\tes$$

This can be a symmetry of the theory provided the quantum numbers  transform
as integers. This places the following constraints on the  parameter  $\rho$
$$\eqalign{\rho{\bf A}  &\in Z,\cr
{\rho\over2} ({\bf A}^t {\bf C}{\bf A} ) &\in Z,\cr
{\rho\over4}({\bf A}^t {\bf C}{\bf A} ) ({\bf A}^t {\bf C}{\bf A} )+({\bf A}^t
{\bf C}{\bf A} )+{1\over\rho}& \in Z,\cr
{\rho\over2} ({\bf A}^t {\bf C}{\bf A} ){\bf A}+{\bf A} &\in Z,\cr
{\rho}{\bf A}{\bf A}^t {\bf C}  &\in Z.}\eqn\nic$$
The discrete Wilson lines  ${\bf A} $ must  satisfy constraints
which are equivelant to preserving world-sheet modular invariance
\footnote*{${\bf Q}^*= {{\bf Q}^t}^{(-1)}$}[\thirty, \twentyseven]
$$\eqalign{ {\bf A}({\bf I} - {\bf Q} ) \, & \in \, Z \cr
{1\over2}{\bf A}^t{\bf CA(I-Q)}+{1\over2}{\bf (I-Q^*)A}^t{\bf
CA} \, & \in \, Z }
\eqn\wsmi$$
For  the orbicircle  \wsmi \ implies that ${\bf A}^t {\bf C} {\bf A} $ is
either zero or half-
integer valued, However the constraints \nic\ can only be solved for
${\bf A} ={\bf  0 }$ and $\rho=1$
which corresponds to the  to the usual
$R\rightarrow1/{2R}$ duality symmetry. This demonstrates that
discrete Wilson lines can
break the stringy duality symmetry.
\chapter{Toroidal Compactification}
In this section, the target space duality symmetry of string compactified on a
$d$-dimensional torus ${\bf T}^d$ [\nineteen,\twenty] is reviewed.
The worldsheet action for the compact coordinates, in the lattice basis, is
given as
$${\cal S}_{torus}={1\over 2\pi}\int d\sigma d\tau
\eta^{\alpha\beta}\Big(G_{ij} \partial_\alpha X^i\partial_\beta
X^j+\epsilon^{\alpha\beta}B_{ij}\partial_\alpha X^i\partial_\beta
X^j\Big)\eqn\act$$
where the metric $G_{ij}$ is defined as the scalar product of the  basis
vectors $e_i$, $i=1, ..., d,$ of the lattice ${\bf\Lambda}_d$ generating the
torus
${\bf T}^d={\textstyle R^d\over\textstyle{\bf\Lambda}_d}$ and $B_{ij}$ is the
antisymmetric tensor.  Together they
describe the $d^2$-dimensional moduli space of toroidal compactification.
Notice that in the action ${\cal S}_{torus}$ the coordinates $X^i$ denote the
components of the internal dimensions in the lattice basis, i.e.,
$X^\mu=e^\mu_iX^i$ where $\mu$ is the  internal space index, therefore on  the
torus
we have $X^i\equiv X^i+2\pi n^i$.
In matrix notation, the left and right moving momenta can be written as
[\twentynine]
$${\bf p}_L={{\bf m}\over 2} +({\bf G}-{\bf B}){\bf n}, \qquad {\bf p}_R={{\bf
m}\over2}- ({\bf G}+{\bf B}){\bf n},\eqn\lrmm$$
where ${\bf n}$ and ${\bf m}$, the windings and the momenta respectively, are
$d$-dimensional integer valued vectors,  while $\bf G$ and $\bf B$
are $d\times d$ matrices representing the background metric and antisymmetric
tensor.
The moduli-dependent part of the scaling dimension and
the spin of the vertex operators are given by
$$\eqalign{H=& {1\over2} ({\bf p}^t_L {\bf G}^{-1} {\bf p}_L + {\bf p}^t_R
{\bf
G}^{-1} {\bf p}_R)={1\over2}  {\bf u}^t \bf\Xi \bf u,\cr
S= &{1\over2} ({\bf p}^t_L {\bf G}^{-1} {\bf p}_L - {\bf p}^t_R {\bf G}^{-1}
{\bf p}_R)={1\over2} {\bf u}^t\bf\eta \bf u,}\eqn\canonical$$
where
$$\bf u = \pmatrix{\bf n \cr \bf m}, \quad \bf\eta = \pmatrix{\bf 0 & {\bf
1}_d
\cr{\bf 1}_d &\bf 0 },\quad \Xi =\pmatrix{2({\bf G}-{\bf B}){\bf G}^{-1}({\bf
G}+{\bf B}) & {\bf B}{\bf G}^{-1}\cr - {\bf G}^{-1} {\bf B}& {1 \over 2} {\bf
G}^{-1}}.\eqn\bmm$$
where ${\bf 1}_d$ is the
identity matrix in $d$ dimensions.

Discrete target space duality symmetries are all those integer-valued linear
transformations of the quantum numbers leaving the spectrum invariant.
Denote these linear transformations by $\bf\Omega$ and write
[\thirty]
$$\bf\Omega: \bf u \longrightarrow {\bf S}_{\Omega}(\bf u) =\bf\Omega^{-1} \bf
u.\eqn\lin$$
To preserve \canonical,  ${\bf\Omega}$ should satisfy the condition
$\bf\Omega^{t}\bf\eta\Omega= \bf\eta$
which  means that ${\bf\Omega}$ is an element of $O(d, d, Z)$. Also, the
moduli
get transformed as
$$\bf\Xi \longrightarrow \bf\Omega^t \bf \Xi \bf\Omega.\eqn\mo$$
The transformation \mo\ contains  the action of the duality group on the
moduli. Obviously,
some $\bf\Omega$ transformations do not mix windings and momenta quantum
numbers, such symmetries are also present in Kaluza-Klein compactification.

The generalization of the above results to the case of orbifolds without
Wilson
lines background is straightforward [\twentynine, \thirty]. The orbifold is
defined as the quotient
of the torus by a group of automorphisms, the point group $P$ of the lattice
which is normally taken to be a cyclic group.
This group has the following action on the quantum numbers  [\twentynine,
\thirty]
$${\bf u} \longrightarrow {\bf u}^\prime = {\cal {\bf R}}{\bf u},\eqn\qu$$
where $\cal {\bf R}$ is given by the matrix

$${\cal {\bf R}}= \pmatrix{{\bf Q} & {\bf 0} \cr{\bf 0}&{\bf Q}^* },\eqn\po$$
and  the matrix $\bf Q$ is defined as
$$\theta ^\mu_\nu e^\nu_i\rightarrow e^\mu_j Q_{ji}\eqn\yet$$
where $\theta ^\mu_\nu$ is the matrix corresponding to the action of the
generator of the point group on the six-dimensional internal space.  For the
point group to be a  lattice
automorphism, the background fields (moduli) must satisfy [\twentyone]
$$ {\bf Q}^t{\bf G}{\bf Q}={\bf G},  \qquad {\bf Q}^t{\bf B}{\bf Q}={\bf
B}.\eqn\co$$

Finally the target space symmetries of the untwisted sector of the orbifold
are those of the torus commuting with the twist matrix ${\cal {\bf R}}.$
More generally, the symmetries are those satisfying ${\bf\Omega} {\cal {\bf
R}}- {\cal {\bf R}}^k{\bf\Omega}={\bf 0},$ $k=1,\dots N$, where $N$ is the
order of the twist [\thirty]. For six-dimensional orbifolds one defines the
twist as
$\theta=({\zeta_1}, {\zeta_2}, {\zeta_3})$
where the notation is such
that the action of $\theta$  in the complex basis is $(e^{2 \pi  i \zeta_1},
e^{2 \pi  i \zeta_2}, e^{2 \pi  i \zeta_3}).$
If the action of the point group generated by $\theta$ has no invariant
planes,
then the twisted sectors of the theory are independent of the moduli. This
means that the twisted states have no windings nor momenta. However, if a
point
group element leaves a particular complex plane invariant then the
corresponding twisted sector states will have non-vanishing windings and
momenta and thus their scaling dimensions will depend on the moduli of the
unrotated plane. Let $\theta^k$ be a group element which leaves a particular
complex plane invariant, then the twisted states winding and momentum will
satisfy

$${\bf Q}^k{\bf n}={\bf n}; \qquad {{\bf Q}^*}^k {\bf m}={\bf m}.\eqn\r $$
The target space symmetries of the twisted sectors for ${\bf Z}_N$ Coxeter
orbifolds have been considered in [\twentyfour].
\chapter{Two-Dimensional Toroidal Compactification.}
In this section, the case of two-dimensional toroidal compactification is
considered. The relevance of this case will become clear in what follows.
It is convenient to group
the four real degrees of freedom of $G_{ij}$ and $B_{ij}=b\epsilon_{ij}$,
parametrizing the background of two-dimensional compactification
into two complex moduli [\seven] defined as
$$\eqalign{T=&T_1+iT_2=2\Big(b+i\sqrt{\det \bf G}\Big),\cr
U=&U_1+iU_2={1\over G_{11}}\Big(G_{12}+i\sqrt{\det \bf G}\Big).}\eqn\shell$$
Clearly, in terms of these moduli, the metric is given by
$${\bf G}={T_2\over 2U_2}\pmatrix{1&U_1\cr U_1&\vert U\vert^2}\eqn\tod$$
and  the moduli-dependent scale and the spin of the vertex operators of the
underlying conformal field theory take the following form
$$\eqalign{H=&{1\over T_2U_2}\vert TUn_2+Tn_1- Um_1+m_2\vert^2\cr S=& m_1n_1+
m_2n_2}\eqn\s$$
where $n_1$, $n_2$, $m_1$ and $m_2$ are integers
 denoting the windings
and momenta.
The spectrum is invariant (in addition to the symmetries $T\leftrightarrow U$,
$T\leftrightarrow -\bar T$, $U\leftrightarrow -\bar U$)
under two copies of the modular group $PSL(2, Z)$ acting on the moduli as
$$U\rightarrow {aU+b\over cU+d},\qquad
T\rightarrow {a'T+b'\over c'T+d'},\eqn\bai$$
where $ad-bc=1$ and $a'd'-b'c'=1.$
These transformations act on the quantum numbers as follows
$$\bf u\rightarrow\pmatrix{{\bf M}& {\bf 0}\cr  {\bf 0}&{({\bf M}^*)}}\bf
u,\quad
\bf u\rightarrow\pmatrix{d'{\bf I}_2&-c'{\bf L}\cr b'{\bf L}&a'{\bf I}_2} \bf
u,\eqn\moh$$
where
$$ {\bf M}=\pmatrix{a&-b\cr -c&d},\quad {\bf L}=\pmatrix {0&1\cr
-1&0}.\eqn\ch$$
In the presence of a ${\bf Z}_2$-twist both moduli $T$ and $U$ are compatible
with the point group and the two-dimensional ${\bf Z}_2$-orbifold has the same
duality symmetries as for the torus.
However for other point groups, only the $T$ moduli survives the twist and the
duality group is given by  $PSL(2, Z)$ acting on $T$ (also $T\leftrightarrow
-\bar T$ is a symmetry of the spectrum) for all two-dimensional ${\bf Z}_N$
orbifold.
\chapter{ Duality
Symmetries  with Discrete Wilson Lines:
${\bf\Lambda}_6={\bf\Lambda}_2+{\bf\Lambda}_4$}
The moduli-dependent threshold corrections to the gauge coupling constants
have
contributions from those twisted sectors  of the orbifold model in which the
point group element (the twist) leave a particular complex plane of the three
complex planes of the orbifold invariant. Those sectors are known as the
${\cal
N}=2$ sectors because they possess two space-time supersymmetries
[\eight,\nine, \ten].
It is the duality symmetries of the spectrum of these sectors which we wish to
study here.
Twisted sectors are only sensitive to the geometry of their invariant planes,
in other words, their states have conformal dimensions depending  on the
moduli $T$ and $U$ associated with the unrotated plane.

In six-dimensional orbifold compactification without Wilson lines and with
lattices decomposable as ${\bf\Lambda}_6=
{\bf\Lambda}_2+{\bf\Lambda}_4$ and where a certain twist leaves a plane
invariant lying
in ${\bf\Lambda}_2$, the corresponding twisted sector have states with windings
and
momenta taking values only in ${\bf\Lambda}_2$ and
its dual ${\bf\Lambda}^*_2.$
The moduli-dependent scaling and spin of these states is thus given as in \s.
Therefore the
spectrum is invariant under the action of $PSL(2, Z)\times PSL(2, Z)$ one for
each complex moduli.

Now in the presence of a quantized Wilson line background,  the twisted states
have the left and right moving momenta,
${\bf P}_L$ and ${\bf P}_R$ given, in the lattice basis, by
[\twentynine, \thirty]
$$\eqalign{{\bf P}_L=&\Big({{\bf m}\over 2} +({\bf G}- {\bf B}-{1\over4}
{\bf A}^t{\bf C}{\bf A}){\bf n}-{1\over2}{\bf A}^t{\bf C}{\bf l},\quad {\bf
l}+{\bf A}{\bf n}\Big),\cr
{\bf P}_R=&\Big({{\bf m}\over2}- ({\bf G}+
 {\bf B}+{1\over4}{\bf A}^t{\bf C}{\bf A}){\bf n}-{1\over2}{\bf A}^t\bf C{\bf
l}, \quad \bf 0\Big),}\eqn\sell$$
where $\bf l$ is the momentum on the $E_8\times E'_8$ lattice, $\bf G$, $\bf
B$
and $\bf A$ are the
metric, antisymmetric tensor and Wilson line background fields, $\bf C$ is the
Cartan metric for the $E_8\times E'_8$ lattice. Let  $\theta$ be the  generator
of the point group of the orbifold and suppose
that the $\theta^k$-twisted sector leave invariant the plane lying in
${\bf\Lambda}_2$. The action of the point group on the quantum numbers can be
written
$${\bf u}\rightarrow {\bf u}'={\cal {\bf R}}{\bf u}\eqn\square$$
where $$\bf u=\pmatrix{\bf n\cr \bf m\cr \bf l}\eqn\al$$
and
$${\cal {\bf R}}=\pmatrix{{\bf Q}&{\bf 0}&\bf 0\cr\bf\alpha&\bf Q^*&(\bf 1-\bf
Q^*){\bf A}^t{\bf C}\cr {\bf A(I-Q)}&\bf 0&\bf I}\eqn\oleg$$
with
$$\alpha={1\over2}{\bf A}^t{\bf CA(I-Q)}+{1\over2}{\bf (I-Q^*)A}^t{\bf
CA}.\eqn\lo$$
Note that the choice of Wilson lines
must satisfy the constraints
  \wsmi\   and hence the entries of ${\cal {\bf R}}$ are all integers
[\thirty].
The states in the $\theta^k$-twisted sector have windings and momenta
satisfying
the condition
$${\cal {\bf R}}^k {\bf u}={\bf u}.\eqn\steve$$
The solution of \steve\ is
$$\eqalign{(\bf{I-Q}^k){\bf n}=&{\bf 0}\cr
(\bf I-{\bf Q^*}^k)({\bf m}-{1\over2}{\bf A}^t{\bf CAn}-{\bf A}^t{\bf
Cl})=&{\bf 0}.}\eqn\pin$$
Assume that  the fixed plane of $\theta^k$ is the first complex plane, then
${\bf Q}^k$ is block diagonal with the $2\times2$ identity matrix as its
leading block. If we consider the new variables $\bf n$ and ${\bf\hat m}={\bf
m}-{1\over2}{\bf A}^t{\bf CAn}-{\bf A}^t{\bf Cl},$ then ${\bf n}$ and
${\bf\hat
 m}$ can only take non-zero values in their first
two components. Define $${\bf u}_{\perp}=\pmatrix{{\bf n}\cr \hat{\bf m}\cr
\hat{\bf l}}\eqn\mh$$
where $\hat{\bf l}=\bf l+\bf An$. This basis diagonalizes the action of the
point group element $\theta$,
$$\theta : \qquad  {\bf u}_{\perp}\rightarrow {\bf u}'_{\perp}={\cal{\bf
R}}_{\perp}{\bf
u}_{\perp},\eqn\george$$ where
$${\cal{\bf R}}_{\perp}=\pmatrix{\bf Q&\bf 0&\bf 0\cr \bf 0&\bf Q^*&0\cr \bf
0&\bf 0&\bf I}.\eqn\blackforest$$

In terms of the new variables, the left and right momenta  ${\bf P}_L$ and
${\bf P}_R$ take the form
$$\eqalign{{\bf P}_L&\equiv\Big({{\hat {\bf m}_0}\over 2} +({\bf
G}_{\perp}-{\bf B}_{\perp}){\bf n}_0,
\quad \hat {\bf l}_0\Big)\cr
{\bf P}_R&\equiv\Big({\hat {\bf m}_0\over 2} -({\bf G}_{\perp}+{\bf
B}_{\perp}){\bf n}_0,\quad
{\bf 0}\Big) \cr
{\hat {\bf m}}_0 \, & = \, {\bf
m}_{0}-{1\over2}{\bf A}^t_{\perp} {\bf C} {\bf A}_{\perp} {\bf n}_{0}-
{\bf A}^t_{\perp} {\bf C}{\bf l} \cr
\hat {\bf l}_0 \, & = \, {\bf l}+ {\bf A}_{\perp} {\bf n}_0 }
\eqn\pat$$
where ${\bf G}_{\perp}$ and ${\bf B}_{\perp}$ are the moduli associated with
the first  complex plane. The quantity ${\bf A}_{\perp}$ is an $8\times 2$
matrix whose elements are the first two columns of $\bf A$ also
${\bf n}_0$ and $\hat{\bf m}_0$ are understood to be 2-dimensional vectors
whose elements are the first two components (the only non-vanishing) of ${\bf
n}$ and $\hat{\bf m}$ respectively .
In this case the scaling dimension of the fields
and their spin is
$$\eqalign{H =&{1\over2} {\bf u}^t_{\perp}{\bf\Xi}_{\perp}{\bf
u}_{\perp}+{1\over2}\hat {\bf l}_0^t{\bf C}\hat {\bf l}_0\cr S=&{1\over2} {\bf
u}^t_{\perp}\eta {\bf u}_{\perp}+{1\over2}
{\hat {\bf l}}_0^t{\bf C}\hat {\bf l}_0.}\eqn\james$$
Here
$${\bf\Xi}_{\perp} =\pmatrix{2({\bf G}_{\perp}-{\bf B}_{\perp}){\bf
G}_{\perp}^{-1}({\bf G}_{\perp}+{\bf B}_{\perp}) & {\bf B}_{\perp}{\bf
G}_{\perp}^{-1}\cr - {\bf G}_{\perp}^{-1} {\bf B}_{\perp}& {1\over 2} {\bf
G}_{\perp}^{-1}},\qquad {\bf u}_{\perp}=\pmatrix{{{\bf n}_0}\cr {\hat {\bf
m}_0}\cr {\hat {\bf l}_0}}.\eqn\mm$$
Thus apart from the additional term coming from the internal gauge lattice,
the
above expressions for $H$ and $S$, when expressed in terms of the complex
moduli $T$ and $U$,  are  identical to those of  \s\  but
with modified quantum numbers.

We define the duality symmetries as those leaving ${\hat{\bf l}}_0$ and the
discrete Wilson lines invariant \footnote*{
It should be remarked that this definition is in contrast to the
situation where continuous Wilson lines are present, since the latter
do transform under duality symmetries [\thirtytwo] .}.
Using the results of the previous section, it  can be easily seen that $H$ and
$S$ \james\  are invariant, in the first instance, under  the transformations
$PSL(2, Z)\times PSL(2, Z)$  acting on
$T$ and $U$ as in \bai, where $T$ and $U$ are constructed out of ${\bf
G}_{\perp}$ and ${\bf B}_{\perp}$
using Eq.\shell.
However there will be a constraint  on the parameters $\{a, b, c, d, a', b',
c', d'\}$
arising from the fact the quantum numbers should transform as integers.
Under the transformations \bai, the modified quantum numbers transform as
follows,
$${\bf u}_{\perp}\rightarrow\pmatrix{{\bf M}& {\bf 0}& {\bf 0}\cr  {\bf
0}&{({\bf M}^*)}&{\bf 0}\cr {\bf 0}&{\bf 0}&{\bf I}}{\bf u}_{\perp},\quad
{\bf u}_{\perp}\rightarrow\pmatrix{d'{\bf I}_2&-c'{\bf L}& {\bf 0}\cr b'{\bf
L}&a'{\bf I}_2&{\bf 0}\cr {\bf 0}&{\bf 0}&{\bf I}}{\bf
u}_{\perp},\eqn\beaver$$
where
${\bf M}$ and  ${\bf L}$ are given by \ch.

In order for the quantum number ${\bf n}_0$, ${\bf m}_0$ and ${\bf l}$ to
transform as integers, the following constraints are obtained
$$\eqalign{{\bf A}_{\perp}({\bf I-M})&\in Z,\cr
{1\over2}{\bf A}_{\perp}^t{\bf C}{\bf A}_{\perp}({\bf I}-{\bf
M})+{1\over2}{\bf
(I-M^*)}{\bf A}_{\perp}^t{\bf C}{\bf A}_{\perp}&\in Z}\eqn\sp$$
for the $U$-duality transformation parameters, and
$$\eqalign{c'{\bf A}_{\perp}&\in Z,\cr
{c'\over2}{\bf A}_{\perp}^t{\bf C}{\bf A}_{\perp}&\in Z,\cr
c'{\bf A}_{\perp}{\bf L}{\bf A}_{\perp}^t{\bf C}&\in Z,\cr
(1-d'){\bf A}_{\perp}-{c'\over2}{\bf A}_{\perp}{\bf L}{\bf A}_{\perp}^t{\bf
C}{\bf A}_{\perp}&\in Z,\cr
(1-a'){\bf C}{\bf A}_{\perp}+{c'\over2}{\bf C}{\bf A}_{\perp}{\bf L}{\bf
A}_{\perp}^t{\bf C}{\bf A}_{\perp}&\in Z,\cr
(1-{a'\over2}-{d'\over2}){\bf A}_{\perp}^t{\bf L}{\bf
A}_{\perp}-{c'\over4}{\bf
A}_{\perp}^t{\bf C}{\bf A}_{\perp}{\bf L}{\bf A}_{\perp}^t{\bf C}{\bf
A}_{\perp}&\in Z.}\eqn\rules$$
For those of the $T$-duality.
The constraints \sp\ and \rules\ therefore
break the $PSL(2, Z)$ duality symmetry associated with
both the complex moduli $T$ and $U$ down to a subgroup.
\chapter{ Duality
Symmetries
with Discrete Wilson Lines: General  ${\bf \Lambda}_6 $.}
In this section we consider the previous discussions for  more general choices
of orbifold lattices ${\bf \Lambda}_6 $.
As an example, consider the Coxeter orbifold
${\bf Z}_6-II-a$, with the twist defined by $\theta=(2,1,-3)/6$ and
${\bf \Lambda}_6 $ given by the
 $SU(6)\times SU(2)$ root lattice. This  model has the first and third planes
unrotated by the $\theta^3$ and $\theta^2$ twists, respectively.
Let us consider the $\theta^2$ twisted sector. We want to determine the
symmetry  group
for the moduli
$(T_3, U_3)$ of
the  third complex plane, which leaves the spectrum of the $\theta^2$ twisted
sectors invariant.

The matrix $\bf Q$ defining the twist action on the quantum numbers is given
by
$${\bf Q}=\pmatrix{0&0&0&0&-1&0\cr
1&0&0&0&-1&0\cr 0&1&0&0&-1&0\cr 0&0&1&0&-1&0\cr 0&0&0&1&-1&0\cr 0&0&0&0&0&-1}
. \eqn\qua$$
The constant background fields compatible with the twist $\theta$ are
given as
$${\bf G}=\pmatrix {r^2&x&l^2&R^2&l^2&u^2\cr
	     x&r^2&x&l^2&R^2&-u^2\cr
	     l^2&x&r^2&x&l^2&u^2\cr
	     R^2&l^2&x&r^2&x&-u^2\cr
	     l^2&R^2&l^2&x&r^2&u^2\cr
	     u^2&-u^2&u^2&-u^2&u^2&y},\eqn\met$$
$${\bf B}=\pmatrix{0&-\beta&-\delta&0&\delta&-\gamma\cr
\beta&0&-\beta&-\delta&0&\gamma\cr \delta&\beta&0&-\beta&-\delta&-\gamma\cr
0&\delta&\beta&0&-\beta&\gamma\cr
-\delta&0&\delta&\beta&0&-\gamma\cr
\gamma&-\gamma&\gamma&-\gamma&\gamma&0},\eqn\su$$
with $R^2=-2l^2-r^2-2x$.

In the $\theta^2$ twisted sector,
 the twisted states are
characterized by the windings and
momenta  satisfying the condition
$$\eqalign{(\bf I-{\bf Q}^2)\bf n=&\bf 0,\cr   (\bf I-{{\bf Q}^*}^2)(\bf
m-{1\over2}\bf A^t\bf C\bf A\bf n-\bf A^t\bf C\bf l)=&\bf 0.}\eqn\kri$$
and are given by
$${\bf n}_{sol}=\pmatrix{n_5\cr 0\cr n_5\cr 0\cr n_5\cr n_6},\qquad \qquad
\qquad
{\hat {\bf m}}_{sol}=\pmatrix{\hat m_5\cr -\hat m_5\cr \hat m_5\cr -\hat
m_5\cr \hat m_5\cr \hat m_6}.\eqn\heb$$
The moduli-dependent scaling and spin $H_3$ and  $S_3$ associated with
the vertex operators creating  the $\theta^2$  twisted states  are given by
$$\eqalign{H=&{1\over2}\tilde{\bf u}_{\perp}^t
\Xi_{\perp}\tilde{\bf u}_{\perp}+
{1\over2}({\bf l}+A_{\perp}{\bf n}_0)^t{\bf C}({\bf l}+{\bf A}_{\perp}{\bf
n}_0),\cr S=&{1\over2} \tilde {\bf
u}_{\perp}^t{\bf\eta}\tilde{\bf u}_{\perp}+{1\over2}({\bf l}+A_{\perp}{\bf
n}_0)^t{\bf C}({\bf l}+{\bf A}_{\perp}{\bf n}_0),}\eqn\ann$$

where $\Xi_{\perp}$ is given by the expression in \mm\ and
$$\tilde{\bf u}_{\perp}=\pmatrix {n_5\cr n_6\cr 3\hat {m}_5\cr \hat {m}_6},
\quad
\quad {\bf G}_{\perp}=\pmatrix{6l^2+3r^2&3u^2\cr 3u^2&y}, \quad
{\bf B}_{\perp}=\pmatrix{0&-\gamma\cr \gamma&0}.$$
$({\bf G}_{\perp}, {\bf B}_{\perp})$ being the metric and the antisymmetric
tensor corresponding to the third complex plane.
Also, ${\bf A}_{\perp}$ is defined by

$${\bf A}{\bf n}_{sol}={\bf A}{\bf K}_2\pmatrix{n_1\cr n_2\cr n_3\cr
n_4\cr n_5 \cr n_6}
\equiv{\bf A}_{\perp}\pmatrix{n_5\cr n_6},\qquad {\bf
K}_2=\pmatrix{0&0&0&0&1&0\cr 0&0&0&0&0&0\cr 0&0&0&0&1&0\cr
0&0&0&0&0&0\cr0&0&0&0&1&0\cr0&0&0&0&0&1}\eqn\ton$$

Note that one can write
$$\pmatrix{{\hat m}_5\cr {\hat m}_6}=
\pmatrix{m_5\cr m_6}-{1\over2} {\tilde{\bf A}^t_{\perp}}{\bf C} {\bf
A}_{\perp}\pmatrix{n_5\cr n_6}- {\tilde{\bf A}^t_{\perp}}{\bf C} {\bf
l},\eqn\ha$$
where
$$
\eqalign{&{\bf K}_3^t\left (\matrix{{\hat m}_1  \cr
\vdots  \cr {\hat m}_6}\right )\equiv \pmatrix{{\hat m}_5\cr {\hat m}_6\cr
0\cr0\cr0\cr0\cr0},\qquad
{\bf K}^t_3=\pmatrix{1&0&0&0&0&0\cr 0&0&0&0&0&1\cr 0&0&0&0&0&0\cr
0&0&0&0&0&0\cr0&0&0&0&0&0\cr 0&0&0&0&0&0}\cr &
{\tilde {\bf A}}^t_{\perp}\equiv\  \hbox{first two rows of}\ {\bf K}_3^t{\bf
A}^t.}\eqn\swan$$
We repeated the analysis for the Coxeter orbifolds  listed in Table $1$ in and
in general
it was found that the moduli-dependent scaling and spin of twisted states with
invariant planes
can always be written in the form
$$\eqalign{H=&{1\over2}\tilde{\bf u}_{\perp}^t
\Xi_{\perp}\tilde{\bf u}_{\perp}+{1\over2}({\bf l}+{\bf A}_{\perp}{\bf
n}_0)^t{\bf C}({\bf
l}+{\bf A}_{\perp}{\bf n}_0), \cr
S=&{1\over2} \tilde{\bf u}_{\perp}^t{\bf\eta}\tilde{\bf u}_{\perp}
+{1\over2}({\bf l}+{\bf A}_{\perp}{\bf n}_0)^t{\bf C}({\bf l}+{\bf
A}_{\perp}{\bf n}_0),}\eqn\nn$$
where
$\tilde{\bf u}_{\perp}=\pmatrix {{\bf n}_0\cr {\bf K}_1\hat {\bf m}_0}$ is a
four-vector containing  the 2-dimensional vector ${\bf n}_0$ and $\hat{\bf
m}_0$, the winding and  momentum of the invariant plane,  ${\bf G}_{\perp}$
and
${\bf B}_{\perp}$ are the metric and antisymmetric tensors corresponding to
the
invariant complex plane and ${\bf K}_1$ is a constant $2\times 2$
integer-valued matrix.
For example, in the
${\bf Z}_6-II-a$ model considered above, the $\theta^2$ sector has
$${\bf K}_1 =\pmatrix{3&0\cr 0&1}.\eqn\worky$$ The values for ${\bf K}_1$ for
some
Coxeter orbifolds are summarized in Appendix A.
Under the duality transformations \bai,
the
quantum numbers $\tilde{\bf u}_{\perp}$ transform as follows,
$$\tilde{\bf u}_{\perp}=\pmatrix{{\bf n}_0\cr {\bf K}_1 \hat{\bf m}_0\cr
\hat{\bf
l}_0}\rightarrow\pmatrix{{\bf M}& {\bf 0}& {\bf 0}\cr  {\bf 0}&{({\bf
M}^*)}& {\bf 0}\cr  {\bf 0}& {\bf 0}&{\bf I}}\tilde{\bf u}_{\perp},\quad
\tilde{\bf u}_{\perp}\rightarrow\pmatrix{d'{\bf I}_2&-c'{\bf L}&0\cr b'{\bf
L}&a'{\bf I}_2& {\bf 0}\cr {\bf 0}& {\bf 0}&{\bf I}}\tilde{\bf
u}_{\perp}.\eqn\san$$
In order for the quantum numbers ${\bf n}_0$, ${\bf m}_0$ and
${\bf
l}$ to transform as integers, the following
constraints must be satisfied,
$$\eqalign{{\bf A}_{\perp}({\bf I}-{\bf M})&\in Z,\cr
{\bf K}_1^{-1}{\bf M}^*{\bf K}_1&\in Z,\cr {\tilde{\bf A}^t_{\perp}}{\bf C}
{\bf
A}_{\perp}-{1\over2}
{\tilde{\bf A}^t_{\perp}}{\bf C} {\bf A}_{\perp}{\bf M}-{1\over2}{\bf
K}_1^{-1}{\bf M}^*{\bf K}_1{\tilde{\bf A}^t_{\perp}}{\bf C} {\bf
A}_{\perp}&\in
Z,\cr
 ({\bf I}-{\bf K}_1^{-1}{\bf M}^*{\bf K}_1){\tilde{\bf A}^t_{\perp}}{\bf
C}&\in
Z,}\eqn\spp$$
and
$$\eqalign{c'{\bf A}_{\perp}{\bf L}{\bf K}_1 &\in Z\cr c'{\bf A}_{\perp}{\bf
L}{\bf
K}_1{\tilde{\bf A}^t_{\perp}}{\bf C}&\in Z,\cr (1-d'){\bf A}_{\perp}
-{c'\over2}{\bf A}_{\perp}
{\bf L}{\bf K}_1{\tilde{\bf A}^t_{\perp}}{\bf C}{\bf A}_{\perp}&\in Z,\cr
c'{\bf
K}_1{\tilde{\bf A}^t_{\perp}}{\bf C}&\in Z\cr {c'\over2}{\bf K}_1{\tilde{\bf
A}^t_{\perp}}{\bf C}{\bf A}_{\perp}&\in Z,\cr {c'\over2}{\tilde{\bf
A}^t_{\perp}}{\bf C}{\bf A}_{\perp}{\bf L}{\bf K}_1&\in Z,\cr
(1-a'-{c'\over2}{\tilde{\bf A}^t_{\perp}}{\bf C} {\bf A}_{\perp}{\bf L}{\bf
K}_1)
{\tilde{\bf A}^t_{\perp}}{\bf C}&\in Z,\cr {\tilde{\bf A}^t_{\perp}}{\bf C}
{\bf A}_{\perp}(1-{a'\over2}-{d'\over2})-{c'\over4}
{\tilde{\bf A}^t_{\perp}}{\bf C} {\bf A}_{\perp}{\bf L}{\bf K}_1
{\tilde{\bf A}^t_{\perp}}{\bf C} {\bf A}_{\perp}+b'{\bf K}_1^{-1}{\bf L}&\in
Z.}
\eqn\morerules$$
The matrices  ${\bf K}_2$ and ${\bf K}_3$ for the Coxeter orbifolds
are listed in  Appendix A.

Note that for the cases when ${\bf K}_1={\beta}{\bf I}$, one
can redefine the $T$  modulus as ${\tilde T}={\textstyle
T\over\textstyle\beta}$
and in terms of the new moduli one obtains the constraints \sp\ and \rules ,
but  with ${{\bf A}^t_{\perp}}$ replaced with ${\tilde{\bf A}^t_{\perp}}$
as defined in \swan\ and with ${\bf A}_{\perp}$ defined by \ton. The
constraints \rules\ don't involve the parameter $b'$ and thus in terms of
the redefined modulus ${\tilde T},$ one obtains a larger symmetry group.
Also, in the absence of Wilson lines but  where ${\bf K}_1 \not={\beta}{\bf
I}$,
the duality group $PSL(2, Z)$ is broken down to a subgroup. This has been
demonstrated in [\twentyfour], and we list in Table 2 these subgroups obtained
for the orbifolds given in Table 1.
As a concrete example, we go back to the  ${\bf Z}_6-II-a$ orbifold and
consider the
following choice of Wilson line,
$${\bf A}^t={1\over2}\pmatrix{0&0&0&0&0&0&0&0\cr 0&0&0&0&0&0&0&0\cr
0&0&0&0&0&0&0&0\cr
0&0&0&0&0&0&0&0\cr0&0&0&0&0&0&0&0\cr 1&0&-1&0&0&0&0&0}.\eqn\ibrahim$$
Using \spp, \morerules, \worky, \ton\ and \swan\  we found that the
$T_3$-duality
symmetry for the $\theta^2$ sector is $\Gamma^0(6)$
and the $U_3$-duality is $\Gamma_0(6).$

For the $\theta^3$ sector,  the value of ${\bf K}_1$ is $2{\bf I}$. In this
case we define a new moduli ${\tilde T}_1={\textstyle {T_1}\over\textstyle 2}$
and employ the constraints \rules\  with  ${\tilde{\bf A}^t_{\perp}}$ and
${{\bf A}_{\perp}}$ as defined by \swan\  and  \ton\ respectively. However with
the above choice of Wilson line, one finds that the values of ${\bf
A}_{\perp}$ and ${\tilde{\bf A}}_{\perp}$ are both vanishing and therefore the
Wilson
line decouple completely from the theory as far as the duality symmetries are
concerned. Thus the duality symmetry for the $\theta^3$ sector in terms of
${\tilde T}_1$ is $PSL(2, Z).$ More examples are listed in Table 3.
In fact, this phenomenon of decoupling of the Wilson lines in  particular
twisted sectors occurs for more general choices of the Wilson lines than
those listed in Table 3. For example, the most general allowed
Wilson lines in the $Z_6-II-b$ orbifold take the form
$${\bf A}^t =\pmatrix{{\bf v}/3 \cr {\bf v}/3 \cr
 {\bf 0}\cr {\bf 0}\cr {\bf w}/2 \cr {\bf w}/2 }  \cdot {\bf X}^{-1}, \quad
{\bf X} ={1\over2} \pmatrix{2&-2&0&0&0&0&0&0\cr
		   0&2&-2&0&0&0&0&0\cr
		   0&0&2&-2&0&0&0&0\cr
		   0&0&0&2&-2&0&0&0\cr
		   0&0&0&0&2&-2&0&0\cr
		   0&0&0&0&0&2&-2&0\cr
		   0&0&0&0&0&0&2&-2\cr
	      -1&-1&-1&-1&-1&1&1&1} \eqn\fin$$
where ${\bf v} $ and ${\bf w} $ are $E_8  $
lattice vectors, (for simplicity we have
only considered Wilson lines in $E_8 $ not $E_8 \times E_8^{'} $ .)
The entry ${\bf 0} $ in
\fin\ denotes a zero vector of the lattice. The corresponding
Wilson lines ${\bf A}_{\perp}$ and $\tilde{\bf A}_{\perp}^t $ in the
$\theta^2 $  twisted sector have the property that they  depend on the
components of the vector ${\bf w} $  only. Similarly, in the $\theta^3 $
sector ${\bf A}_{\perp}$ and $\tilde{\bf A}_{\perp}^t $ depend
on the
components of ${\bf v} $ only. Thus if we choose any  Wilson line
with ${\bf v} \, = \, 0 \, , \, {\bf w} \neq 0 $ we would have decoupling
in the $\theta^3 $ sector, and hence a $PSL(2, Z) $ duality group. A similar
phenomenon occurs for all the ${\bf Z}_6 $ orbifolds.
 When decoupling occurs in a twisted
sector, one can explicitly compute the moduli dependent
threshold correction coming from this sector, despite
the fact that the Wilson lines do break the gauge symmetry.

In summary, we have studied the duality symmetries of orbifolds moduli
associated with invariant planes. These are  relevant to  threshold
corrections
to the gauge coupling constants in the presence of discrete Wilson lines,
which
are not yet known.
The automorphic functions of the subgroups obtained should be related to the
expression of these corrections. The threshold corrections are essential
in the evaluation of any possible non-perturbative superpotential
in addition to their relevance to gauge couplings unification. They are also of
fundamental importance in the study of supersymmetry breaking.
Our methods should be relevant to the study of duality symmetries in the
presence of continous Wilson
lines [\thirtytwo, \thirtythree].  We hope to report on these questions  in
future publications.

\centerline{\bf{ACKNOWLEDGEMENT}}
S.T. would like to thank H. P. Nilles and S. Steiberger for useful
conversations.
This work is supported in part by S.E.R.C. and the work of S.T  is
supported by the Royal Society.
\vfill\eject
\centerline{\bf {Table Captions}}
\noindent
{\bf Table}. 1. Non-decomposable $Z_N$ orbifolds. For the point
group generator $\theta $, we display $ ( \zeta_1 , \zeta_2 , \zeta_3 ) $
such that the action of $\theta $ in the complex plane  orthogonal
basis is $(e ^{2\pi i \zeta_1}, e ^{2\pi i \zeta_2},e ^{2\pi i \zeta_3} ). $
The twisted sectors displayed are those with invariant complex planes.
\vskip0.2cm
\noindent
{\bf Table}. 2. Duality groups for the twisted sectors displayed in Table 1.
$T_i$ and $U_i$ are the complex moduli associated with the invariant $i$-$th$
plane.
\vskip0.2cm
\noindent
{\bf Table}. 3. This table contains explicit examples of duality groups for
non-decomposable ${\bf Z}_N$ Coxeter orbifolds, the second column in the table
displays the non-vanishing components of a particular choice of Wilson line.
Note that we have not included $Z_{12}-I-a$, as no Wilson lines are allowed in
this example.
\vskip0.5cm
\vfill\eject
\centerline {{\bf TABLE} 1}
\vskip 1cm
\begintable
Orbifold |Point group generator|Twisted sector|Lattice\cr
$Z_4-a$ | $(1,1,-2)/4$|$\theta^2$ |$SU(4)\times SU(4)$
\cr
$Z_4-b$ | $(1,1,-2)/4$|$\theta^2$ |$SU(4)\times SO(5)\times SU(2)$
\cr
$Z_6-II-a$ | $(2,1,-3)/6$|$\theta^2$,  $\theta^3$|$SU(6)\times SU(2).$ \cr
$Z_6-II-b$ | $(2,1,-3)/6$| $\theta^2$,  $\theta^3$|$SU(3)\times SO(8).$ \cr
$Z_6-II-c$ | $(2,1,-3)/6$| $\theta^2$,  $\theta^3$|$SU(3)\times SO(7)\times
SU(2).$ \cr
$Z_8-II-a$ | $(1,3,-4)/8$|$\theta^2$|$SU(2)\times SO(10)$ \cr
$Z_{12}-I-a$| $(1,-5,4)/12$|$\theta^3$|$E_6$
\endtable
\vskip 1cm
\centerline {{\bf TABLE} 2}
\vskip 1cm
\begintable
Orbifold| Duality group\cr
$Z_4-a$| $\Gamma_{T_3/2}$=$\Gamma^0(2),$\quad
$\Gamma_{U_3}$=$PSL(2,Z)$\cr
$Z_4-b$| $\Gamma_{T_3}$=$\Gamma^0(2),$\quad
$\Gamma_{U_3}$=$\Gamma_0(2).$\cr
$Z_6-II -a$|$\Gamma_{T_3}$=$\Gamma^0(3)$,\quad
$\Gamma_{U_3}=\Gamma_0(3)$, \quad$\Gamma_{T_1/2}$=$PSL(2, Z)$\cr
$Z_6-II -b$|$\Gamma_{T_3}$=$\Gamma^0(3)$,\quad
$\Gamma_{(U_3+2)}=\Gamma^0(3)$, \quad $\Gamma_{T_1}$=$PSL(2, Z)$\cr
$Z_6-II-c$|$\Gamma_{T_3}$=$\Gamma^0(3)$,\quad $\Gamma_{U_3}=\Gamma_0(3)$, \quad
$\Gamma_{T_1}$=$PSL(2, Z)$\cr
$Z_8-II -a$|$\Gamma_{T_3}$=$\Gamma^0(2)$,\quad
$\Gamma_{U_3}=\Gamma_0(2)$\cr
$Z_{12}-I-a$|$\Gamma_{T_3/2}$=$PSL(2, Z)$
\endtable
\vfill\eject
\centerline {{\bf TABLE} 3}
\vskip 1cm
\begintable
Orbifold |Non-zero components of $\bf A$|Duality Symmetry\cr
$Z_4-a$|$A_{11}=-A_{31}=$|$T_3$: $\Gamma^0(2)$,\nr
|$A_{12}=-A_{32}=$ | $U_3$: $\Gamma^0(2)$\nr
|$A_{13}=-A_{33}={1\over2}$|
\cr
$Z_4-b$|$A_{11}=-A_{31}=$|$T_3$: $\Gamma^0(2)$, \nr
|$A_{12}=-A_{32}=$  |$U_3$: $\Gamma_0(2)$\nr

|$A_{13}=-A_{33}={1\over2}$|
\cr
$Z_6-II-a$ |$A_{16}=-A_{36}={1\over2}$|$T_3$:
$\Gamma^0(6)$,   \nr |    |$U_3$: $\Gamma_0(6)$\nr  | | ${T_1/2}$: $PSL(2,
Z)$\cr
$Z_6-II-b$ |$A_{15}=-A_{35}=$|$T_3$:
$c=0\  (mod\ 2), \ b= 0\  (mod \ 3)$,   \nr | $A_{16}=-A_{36}={1\over2}$
|$U_3+2$: $\Gamma^0(6)$\nr  | | $T_1$: $PSL(2, Z)$\cr
$Z_6-II-c$ |$A_{16}=-A_{36}={1\over2}$|
$T_3$: $c=0\  (mod\ 2), \ b= 0\  (mod \ 3),$ \nr | |$U_3$: $\Gamma_0(6)$\nr |
|$T_1$: $PSL(2, Z)$\cr
$Z_8-II-a$|$A_{14}=-A_{34}=$|
$T_3$:  $c=0\  (mod\ 2), \ b= 0\  (mod \ 2)$, \nr
|$A_{15}=-A_{35}=$ | $U_3$: $\Gamma_0(4)$\nr
|$A_{16}=-A_{36}={1\over2}$|
\endtable
\vskip 0.5cm
\vfill\eject
\Appendix A
$\underline{Z_4-a}$

The $\theta^2$ sector
$${\bf K}_1=\pmatrix{2&0\cr 0&2}$$
$${\bf K}_2=\pmatrix{1&0&0&0&0&0\cr 0&0&0&0&0&0\cr 1&0&0&0&0&0\cr
0&0&0&1&0&0\cr0&0&0&0&0&0\cr 0&0&0&1&0&0},\quad
 {\bf K}^t_3=\pmatrix{1&0&0&0&0&0\cr 0&0&0&1&0&0\cr 0&0&0&0&0&0\cr
0&0&0&0&0&0\cr0&0&0&0&0&0\cr 0&0&0&0&0&0}.$$

$\underline{Z_4-b}$

The $\theta^2$ sector

$${\bf K}_1=\pmatrix{2&0\cr 0&1}$$
$${\bf K}_2=\pmatrix{1&0&0&0&0&0\cr 0&0&0&0&0&0\cr 1&0&0&0&0&0\cr
0&0&0&0&0&0\cr0&0&0&0&0&0\cr0&0&0&0&0&1},\quad
 {\bf K}^t_3=\pmatrix{1&0&0&0&0&0\cr 0&0&0&0&0&1\cr 0&0&0&0&0&0\cr
0&0&0&0&0&0\cr0&0&0&0&0&0\cr 0&0&0&0&0&0}.$$

$\underline{Z_6-II-a}$

The $\theta^2$ sector
$${\bf K}_1=\pmatrix{3&0\cr 0&1}$$
$${\bf K}_2=\pmatrix{0&0&0&0&1&0\cr 0&0&0&0&0&0\cr 0&0&0&0&1&0\cr
0&0&0&0&0&0\cr0&0&0&0&1&0\cr0&0&0&0&0&1}\quad
{\bf K}^t_3=\pmatrix{0&0&0&0&1&0\cr 0&0&0&0&0&1\cr 0&0&0&0&0&0\cr
0&0&0&0&0&0\cr0&0&0&0&0&0\cr 0&0&0&0&0&0}.$$

The $\theta^3$ sector
$${\bf K}_1=\pmatrix{2&0\cr 0&2}$$
$${\bf K}_2=\pmatrix{1&0&0&0&0&0\cr 0&1&0&0&0&0\cr 0&0&0&0&0&0\cr
1&0&0&0&0&0\cr0&1&0&0&0&0\cr0&0&0&0&0&0}\quad
{\bf K}^t_3=\pmatrix{1&0&0&0&0&0\cr 0&1&0&0&0&0\cr 0&0&0&0&0&0\cr
0&0&0&0&0&0\cr0&0&0&0&0&0\cr 0&0&0&0&0&0}.$$

$\underline{Z_6-II-b}$

The $\theta^2$ sector
$${\bf K}_1=\pmatrix{1&1\cr 1&-2}$$
$${\bf K}_2=\pmatrix{0&0&0&0&0&0\cr 0&0&0&0&0&0\cr 0&0&1&0&0&0\cr
0&0&0&0&0&0\cr0&0&0&0&1&0\cr 0&0&1&0&-1&0},\quad
 {\bf K}^t_3=\pmatrix{0&0&1&0&0&0\cr 0&0&0&0&1&0\cr 0&0&0&0&0&0\cr
0&0&0&0&0&0\cr0&0&0&0&0&0\cr 0&0&0&0&0&0}.$$

The $\theta^3$ sector
$${\bf K}_1=\pmatrix{1&0\cr 0&1}$$
$${\bf K}_2=\pmatrix{1&0&0&0&0&0\cr 0&1&0&0&0&0\cr 0&0&0&0&0&0\cr
0&0&0&0&0&0\cr0&0&0&0&0&0\cr 0&0&0&0&0&0}\, = \quad
 {\bf K}^t_3$$

$\underline{Z_6-II-c}$

The $\theta^2$ sector
$${\bf K}_1=\pmatrix{3&0\cr 0&1},$$
$${\bf K}_2=\pmatrix{0&0&0&0&0&0\cr 0&0&0&0&0&0\cr 0&0&1&0&0&0\cr
0&0&0&0&0&0\cr0&0&1&0&0&0\cr0&0&0&0&0&1} \quad
{\bf K}^t_3=\pmatrix{0&0&0&0&1&0\cr 0&0&0&0&0&1\cr 0&0&0&0&0&0\cr
0&0&0&0&0&0\cr0&0&0&0&0&0\cr0&0&0&0&0&0}$$

The $\theta^3$ sector
$${\bf K}_1=\pmatrix{1&0\cr 0&1}$$
$${\bf K}_2=\pmatrix{1&0&0&0&0&0\cr 0&1&0&0&0&0\cr 0&0&0&0&0&0\cr
0&0&0&0&0&0\cr0&0&0&0&0&0\cr 0&0&0&0&0&0}\,= {\bf K}^t_3.$$

$\underline{Z_8-II-a}$

The $\theta^2$ sector
$${\bf K}_1=\pmatrix{2&0\cr 0&1}$$
$${\bf K}_2=\pmatrix{0&0&0&0&0&0\cr 0&0&0&0&0&0\cr 0&0&0&0&0&0\cr
0&0&0&1&0&0\cr0&0&0&-1&0&0\cr0&0&0&0&0&1}, \quad
 {\bf K}^t_3=\pmatrix{0&0&0&1&0&0\cr 0&0&0&0&0&1\cr 0&0&0&0&0&0\cr
0&0&0&0&0&0\cr0&0&0&0&0&0\cr0&0&0&0&0&0}$$
\Appendix B
In this appendix we demonstrate that the modular subgroups obtained
as duality symmetries listed in Table 3, are independent
of the particular realization we choose for the windings and momenta
that lie in an ${\cal N} = 2$ fixed plane. To illustrate
this independence, we shall consider a particular example, namely
$Z_6 -II-b$ in the $\theta^2 $ twisted sector; generalization
to all other cases will then become apparent. We may represent the
windings and momenta
for ${\bf n}$ and $\hat{\bf m}$  of states in the  $\theta^2 $ twisted sector
in
the following ways,
$$\eqalign{ {\bf n}_{sol} \, =& \, \pmatrix{0\cr0\cr{ n}_5 + { n}_6 \cr
0\cr{ n}_5 \cr { n}_6}
 \, = \, \pmatrix{0\cr0\cr{ n}_3 \cr
  0\cr{ n}_3 - { n}_6  \cr { n}_6} \cr
  &\cr
\hat{\bf m}_{sol} \, = & \, \pmatrix{0\cr0\cr \hat{ m}_5 + \hat{ m}_6 \cr
-\hat{ m}_5 - \hat{ m}_6\cr  \hat{ m}_5 \cr \hat{ m}_6 } \, = \,
\pmatrix{0\cr0\cr \hat{ m}_3 \cr
-\hat{ m}_3\cr \hat{ m}_3 -  \hat{ m}_6 \cr \hat{ m}_6} \, = \,
 \pmatrix{0\cr0\cr -\hat{ m}_4 \cr  \hat{ m}_4 \cr
 -\hat{ m}_4 -  \hat{ m}_6 \cr \hat{ m}_6 }  } \eqn\apone$$
${\bf n}_{sol}$ and $\hat{\bf m}_{sol}$ are solutions to the constraints
\steve\ with $k \, = \, 2$, and
any pair of choices for ${\bf n}_{sol} $ and $\hat{\bf m}_{sol} $ should
 give rise to the same modular subgroup. To prove this consider
 a definite choice for ${\bf n}_{sol} $ and $\hat{\bf m}_{sol}$ $e.g$.,
$$\eqalign{ {\bf n}_{sol} \, =& \, \pmatrix{0\cr0\cr{ n}_5 + { n}_6 \cr
0\cr{ n}_5 \cr { n}_6} \, =  {\bf K}_2\, \pmatrix{n_1 \cr n_2\cr n_3\cr
n_4 \cr n_5\cr n_6} \cr
&\cr
\hat{\bf m}_{sol} \, = & \, \pmatrix{0\cr0\cr \hat{ m}_5 + \hat{ m}_6 \cr
-\hat{ m}_5 - \hat{ m}_6\cr  \hat{ m}_5 \cr \hat{ m}_6 }.} \eqn\aptwo$$
Because of the choice of $\hat{\bf m}_{sol} $ in \aptwo, the
corresponding projection  matrix
${\bf K}_3^t $ which was defined earlier in the text, is given by
$$  {\bf K}_3^t \, \pmatrix{\hat{m}_1 \cr \hat{m}_2 \cr \hat{m}_3 \cr
\hat{m}_4 \cr \hat{m}_5 \cr \hat{m}_6 }\, = \, \pmatrix{\hat{m}_5\cr
\hat{m}_6 \cr 0\cr 0\cr 0\cr 0}.\eqn\apthree$$
The modular subgroups corresponding to this choice of solutions will be
determined by the `effective' 2-dimensional Wilson lines ${\bf A}_{\perp} $
and $\tilde{\bf A}_{\perp}^t $ defined in terms of the 6-dimensional Wilson
line
and the matrices ${\bf K}_2 $ and ${\bf K}^t_3 $. Now consider a different
choice of solutions given in \apone, $e.g$.,
$${{\bf n}'}_{sol} \, =\pmatrix{0\cr0\cr{ n}_3 \cr
0\cr{ n}_3-n_6 \cr { n}_6};\qquad
\hat{{\bf m}'}_{sol} \, =\pmatrix{0\cr0\cr \hat{ m}_3 \cr
-\hat{ m}_3\cr  \hat{ m}_3 - \hat{m}_6 \cr \hat{ m}_6 } \eqn\apfour$$
Correspondingly, there will be new matrices ${\bf K}_2 \, ,\, {\bf K}_3^t ,$
${{\bf A}'}_{\perp}$ and $\tilde{{\bf A}'}_{\perp}^{t}$. It is
straightforward to determine the connection between these two sets
of Wilson lines,
$$ {\bf A}_{\perp} \, = \, {{\bf A}'}_{\perp} \, {\cal F} \quad
\tilde{{\bf A}'}_{\perp }^{t} \, = \, {\cal F} \, \tilde{\bf
A}_{\perp}^t\eqn\apfive $$
where in \apfive, ${\cal F}$ is
$${\cal F} \, = \, \pmatrix{1&1\cr0&1}. \eqn\apsix$$
In general we would have found two different matrices ${\cal G} $ and
${\cal F}$ connecting  ${\bf A}_{\perp} \, ,\tilde{{\bf A}}_{\perp}^{t}$ to
the same quantities primed. With the choice of solutions in \aptwo\ and
\apfour, ${\cal G}\, = \, {\cal F} $.

We have to show now, that the constraints on $T$ and $U$ duality
due to ${\bf A}_{\perp} \, ,  \tilde{\bf A}_{\perp}^{t} $ are the same
as those due to ${{\bf A}'}_{\perp} \, ,  \tilde{{\bf A}'}_{\perp}^{t}$
leading to the same  modular subgroup. In the basis defined by
 ${\bf n}_{0} \, = \, \pmatrix{n_5\cr n_6}, \, \hat{\bf m}_{0} \, =
 \, \pmatrix{\hat{m}_5 \cr \hat{m}_6 }$ the  transformations
 under $T$ and $U$ duality symmetries are
 $$\eqalign{ {\bf n}_{0} \, &\rightarrow \, d' {\bf n}_0 - c' \, {\bf L}\, {\bf
K}_1
 \hat{\bf m}_0  \, , \quad {\bf K}_1 \hat{\bf m}  \, \rightarrow \,
 b' {\bf L}\, {\bf n}_0 + a' {\bf K}_1 \, \hat{\bf m}_0 \cr
 {\bf n}_0 \, &\rightarrow \, {\bf M} {\bf n}_0 \, , \quad
 {\bf K}_1 \hat{\bf m}_0 \, \rightarrow \, {\bf M}^* \, {\bf K}_1 \hat{\bf m}_0
}
 \eqn\apseven $$
 In the new basis ${\bf n}'_0 \, = \pmatrix{n_3 \cr n_6},
 \hat{\bf m}'_0 \, = \, \pmatrix{\hat{m}_3 \cr \hat{m}_6 } $
 the same transformations become
 $$\eqalign{  {\bf n}'_0 \, & \rightarrow \, d' {\bf n}'_{0} -
 c' {\cal F} \, {\bf L}\, {\bf K}'_1
 \hat{\bf m}'_{0} \cr
 {\bf K}'_{1}\hat{\bf m}'_0 \, &\rightarrow \,
 b' {\bf L}\, {\cal{F}}^{-1}{\bf n}'_0 + a' {\bf K}'_1 \, \hat{\bf m}'_0 \cr
 {\bf n}'_0\, &\rightarrow \, {\cal{F}} {\bf M} {\cal{F}}^{-1}  {\bf n}'_0\cr
 {\bf K}'_1\hat{\bf m}'_0 \, &\rightarrow \, {\bf M}^* \, {\bf K}'_1
 \hat{\bf m}'_0 .} \eqn\apeight $$
It is now straightforward to derive the  constraints on $T$ and $U$
duality transformations in this new basis, with the Wilson lines
${\bf A}'_{\perp} \,$and $  \tilde{{\bf A}'}_{\perp}^t$ following the
procedures  outlined in section 6. As regards the constraints on
$T$ and $U$ duality, one finds that  conditions  \spp\ and \morerules\  are
reproduced in terms of the
Wilson lines ${\bf A}_{\perp} \, $ and $  \tilde{\bf A}_{\perp}^{t} $
 except that  in various equations there is pre  andways or post
 multiplication by either
${\cal F}^{-1} $, or ${\cal F}.$
However, since ${\cal F}^{-1} \, =\, \pmatrix{1&-1\cr
0&1}$ is itself integer-valued, it is clear that if constraints
\spp\ and \morerules\ are satisfied in the `old' basis, they will remain so in
the new.
Hence, the modular subgroups that satisfy these conditions are independent
of the two choices \aptwo\ and \apfour\ for ${\bf n}_{sol} \, $ and
$\hat{\bf m}_{sol} $. Furthermore, it is now apparent that for ${\it any}$
pair of solutions  ${\bf n}_{sol} \, $and $ \hat{\bf m}_{sol} $
for the $Z_6 - II-b$ orbifold  given
in \apone\  we will arrive at the same conclusion, because the
corresponding matrices ${\cal F},\,$ and ${\cal G}\,$, together with
their inverses, are always  integer-valued. Generalization to all other
Coxeter orbifolds follows directly. Since in any given
orbifold ${\bf n}_{sol}$ and $\hat{\bf m}_{sol} $ satisfy the
${\cal N}  = 2$ constraints \r\ involving the
integer-valued matrix ${\bf Q} $ and
its powers, the corresponding matrices ${\cal F}$ and ${\cal G} $ will
inevitably be integer-valued. That the same is true for   ${\cal F}^{-1}$
and ${\cal G}^{-1} $ follows from the fact that ${\cal{\bf Q}}^{-1} $
is also integer-valued, since ${\theta }^{-1} $ is a lattice automorphism.

\refout
\end